\documentstyle[namedreferences,psfig,solphys]{spackapp}
\def\verbfont{\small\tt}

\newcommand{\citeN}[1]{\citeauthor{#1}\ (\citeyear{#1})}
\newcommand{\citeNP}[1]{\citeauthor{#1},\ \citeyear{#1}}

\newcommand{\ujlisti}{
\itemsep=0 em
\parsep=0.5 em
\partopsep=0.25 em
\topsep=0 em}
\newcommand{\ujlistii}{
\itemsep=0 em
\parsep=0.5 em
\partopsep=0.25 em
\topsep=0 cm}

\newenvironment{lista}{\begin{list}{--}{\ujlisti}}{\end{list}}

\renewcommand{\[}{\begin{equation}}
\renewcommand{\]}{\end{equation}}
\newcommand{\scri}{\scriptsize} 
\newcommand{\vc}[1]{\mbox{\bf #1}}

\newcommand{\dfrac}[2]{\displaystyle\frac{#1}{#2}}

\newcommand{\ov}{\overline}

\newcommand{\pdv}[2]{\frac{\partial #1}{\partial #2}}

\newcommand{\dv}[2]{\frac{{\rm d}#1}{{\rm d}#2}}

\def\rvar{\tilde r}

\def\defdef{\buildrel \rm def \over =}
\def\la{\mathrel{\mathchoice {\vcenter{\offinterlineskip\halign{\hfil
 $\displaystyle##$\hfil\cr<\cr\sim\cr}}}
 {\vcenter{\offinterlineskip\halign{\hfil$\textstyle##$\hfil\cr
 <\cr\sim\cr}}}
 {\vcenter{\offinterlineskip\halign{\hfil$\scriptstyle##$\hfil\cr
 <\cr\sim\cr}}}
 {\vcenter{\offinterlineskip\halign{\hfil$\scriptscriptstyle##$\hfil\cr
 <\cr\sim\cr}}}}}
\def\ga{\mathrel{\mathchoice {\vcenter{\offinterlineskip\halign{\hfil
 $\displaystyle##$\hfil\cr>\cr\sim\cr}}}
 {\vcenter{\offinterlineskip\halign{\hfil$\textstyle##$\hfil\cr
 >\cr\sim\cr}}}
 {\vcenter{\offinterlineskip\halign{\hfil$\scriptstyle##$\hfil\cr
 >\cr\sim\cr}}}
 {\vcenter{\offinterlineskip\halign{\hfil$\scriptscriptstyle##$\hfil\cr
 >\cr\sim\cr}}}}}
 
\def\DGW{{\ov D}_{\mbox{\scri GW}}}

\runningtitle{Making Sense of Sunspot Decay, I}
\runningauthor{Petrovay and van~Driel-Gesztelyi}

\begin{opening}

\title{MAKING SENSE OF SUNSPOT DECAY}
\subtitle{I: Parabolic Decay Law and Gnevyshev--Waldmeier Relation}                                         

\author{K. \surname{Petrovay}}
\institute{Instituto de {Astrof\'\i sica} de Canarias, 
	   La Laguna, Tenerife, E-38200 Spain} 
\author{L. \surname{van~Driel-Gesztelyi} } %\thanks{%
\institute{Observatoire de Paris, 
	   DASOP, F-92195 Meudon Cedex, France, and\\
	   Konkoly Observatory, Budapest, Pf. 67, H-1525, Hungary}
\end{opening}

\begin{ao}
\\
K. Petrovay\\
Instituto de Astrof\'\i sica de Canarias\\
La Laguna, Tenerife, E-38200 Spain\\
E-mail: kpetro@iac.es
\end{ao}                   

\begin{document}

\begin{abstract}
In a statistical study of the decay of individual sunspots based on DPR data
we find that the mean instantaneous area decay rate is
related to the spot radius $r$ and the maximal radius $r_0$ as
$D=C_D \,r/r_0$, $C_D=32.0\pm 0.26\,$MSH/day. 
This implies that sunspots on the mean follow a parabolic decay law; the
traditional linear decay law is excluded by the data. The validity of the 
Gnevyshev--Waldmeier relationship between the maximal area $A_0$ and lifetime 
$T$ of a spot group, $A_0/T\simeq 10\,$MSH/day, 
is also demonstrated for individual sunspots. No evidence is found
for a supposed supergranular ``quantization'' of sunspot areas. 
Our results strongly support the recent turbulent erosion model of sunspot
decay while all other models are excluded. 
\end{abstract}

%\motto{This is a motto\\
%      For details see section~\ref{motto}}
  
%\keywords{LaTeX, Kluwer Style File, Authors' Instructions}

\abbreviations{CM -- central meridian; DPR -- Debrecen Photoheliographic 
	       Results; GPR -- Greenwich 
	       Photoheliographic Results; MSH -- millionth solar hemisphere; 
	       MSHER -- MSH equivalent radius}

%\classification{JEL codes}{D24, L60, 047}

\section{Introduction}
It has been known since the era of the first telescopic observations that
sunspots disappear from the face of the Sun on a timescale of days/weeks by a
gradual shrinking (sometimes punctuated by fragmentation events). During this 
process of decay the boundary of the spot remains well defined and its 
temperature deficit does not change too much (nor does its magnetic field
strength vary by more than about 20\%, cf.\ \citeNP{Collados+};
\citeNP{Steinegger+:balance}). The extremely 
large individual variations
among spots however did not make it possible to find quantitative 
regularities in the decay until the availability of sufficiently large and 
precise observational databases (GPR, Pulkovo, Mt.~Wilson). These studies 
began in earnest towards the middle of this century, and 
by about 1970, two simple statistical relationships had become rather widely
accepted among solar physicists:
\begin{lista}
\item The rule of the proportionality of the maximal area $A_0$ of a sunspot 
group to its lifetime $T$ (first plotted by \citeNP{Gnevyshev}, and formulated 
by \citeNP{Waldmeier:book}):
\[ A_0= \DGW T   \qquad  \DGW \sim 10\,\mbox{MSH/day} \label{eq:GW} .\]
%(1 MSH= one millionth solar hemisphere.) 

\item The alleged linear time-dependence of the area during the phase of decay:
\[ A=A_0 - \overline D (t-t_0 ). \] 
This formula was originally proposed by \citeN{Bumba:decay} for recurrent 
sunspot groups, while for the majority of non-recurrent groups a concave decay
curve was found to be a better fit. However, for a variety of reasons 
including bad sampling intervals and large scatter, 
significant departures of the decay curve from linear were difficult to find
and even more difficult to describe quantitatively. Thus, the linear decay law
had become widely accepted as the simplest approximation to the decay curves
(\citeNP{Ringnes:shortspots}, \citeyear{Ringnes:longspots};
\citeNP{Robinson+Boice:decay}).
\end{lista}

\smallskip

More recently, doubt has been cast on the validity of the linear decay 
law. \citeN{FMI+MVA} and \citeN{VMP+:periph.decay} pointed out that the 
statistical evidence for a parabolic decay law 
\[ A=A_0-2\sqrt{\pi A_0}w(t-t_0) +\pi w^2(t-t_0)^2   \label{parabdec} \]
is somewhat stronger than for a linear decay.
Such a decay law would follow from a slow inwards motion of the spot boundary 
with a constant velocity $w$. However, a definite conclusion about the 
significance of that result could not be drawn. The correct form of the sunspot 
decay law has thus remained undetermined.

Note also that the Gnevyshev--Waldmeier relation (1) was derived for sunspot 
\it groups, \rm 
and its validity for individual spots has not been demonstrated. 

Another open question is that of the existence of preferred length/area scales
corresponding to higher stability of the spots (i.e.\ to lower decay rates). 
Such a ``quantized'' character of the distribution of sunspot areas (possibly
caused by interaction with supergranulation) has been 
proposed by several authors (\citeNP{Ikhsanov}; \citeNP{Dmitrieva+}; 
\citeNP{Bumba+:quantized}). The statistical significance of those findings 
however remains dubious. 

In this paper we present the results of a statistical investigation 
aimed at resolving these fundamental open questions related to sunspot decay.
(Preliminary results of this study have been published elsewhere: 
\citeNP{Petrovay+vDG:ASPE}.)
Our investigation yields definitive answers to the issues mentioned above. 
This has been made possible by our use of the Debrecen Photoheliographic 
Results: this recently published database is the only data set of statistically 
significant size to contain parameters for individual sunspots with day-to-day 
identification, measured and reduced by very high standards of precision. 

The available theories and models of sunspot decay served as our guide during 
the analysis by concentrating our attention to those correlations 
for which different models had different predictions. We are thus also able to 
exclude some of those models, leaving a single one compatible with the 
observational constraints. The models and their predictions are outlined in 
Section 2. In Section 3 we discuss our selection and handling of data in 
detail, while Sections 4 and 5 present the results. Section 6 concludes the 
paper.

\section{Models}
The first quantitative model for sunspot decay was proposed by 
\citeN{Gokhale+Zwaan}. To explain the observed sharp boundary of spots they 
assumed that the magnetic flux tube is surrounded by a current sheet, i.e.\ 
that moving outwards from the spot center, the magnetic flux density drops to 
zero in a very narrow radius range. They further assumed that turbulence is 
fully inhibited by the strong magnetic field in and inside the current sheet, 
so that the decay of the spot is only due to pure Ohmic dissipation in the 
current sheet. They pointed out that a linear decay law implies that the 
thickness of the current sheet must be a constant fraction of the radius, 
i.e.\ the spot remains \it self-similar\/ \rm during the decay. In order to 
produce an ever thinner current sheet, however, they needed to postulate \it ad 
hoc\/ \rm the existence of a radial inflow towards the tube, for which no 
observational indication existed. For this reason, their model never became so 
widely accepted as the turbulent diffusion model.

The latter model, proposed independently by \citeN{Meyer+} and by 
\citeN{Krause+Rudiger}, assumes that the decay of the spot is caused by 
turbulent diffusion throughout the whole cross-section of the tube, with a 
magnetically reduced but otherwise constant diffusivity. It was found that 
defining the boundary of the spot at some arbitrary fixed value of the flux 
density, such a diffusion naturally results in a linear decay law, and the 
Gnevyshev--Waldmeier relation is also returned. The diffusivity was essentially 
treated as an adjustable parameter. The tubes in this model had no sharp 
boundary, being gradually ``washed away'' by diffusion, and the central flux 
density showed a gradual decrease instead of remaining more or less constant 
until the full decay of the spot. The model is also problematic from a 
theoretical point of view, as a constant magnetically reduced diffusivity is 
hardly compatible with a magnetic field that must obviously be lower outside 
the tube. Despite these obvious shortcomings, its prediction of the two 
(alleged) linear laws of sunspot decay made this model the most accepted 
theoretical explanation of the decay for two decades. 

The studies of \citeN{FMI+MVA} and \citeN{VMP+:periph.decay} questioned 
the validity of the linear decay law, reviving the old suggestion 
(\citeNP{Simon+Leighton:supgr.obs}) that the decay of a spot may proceed by the eroding 
action of \it external\/ \rm turbulence that ``gnaws'' off bits and pieces of 
the flux tube. It was time to reconsider the quantitative decay models. The 
recent \it turbulent erosion model\/ \rm (\citeNP{Petrovay+FMI:erosion}) is 
formally a 
generalization of the turbulent diffusion model allowing a variable diffusivity 
explicitly depending on the magnetic field. The solutions of the 
nonlinear diffusion equation in cylindrical geometry
\begin{equation} 
  \frac{\partial}{\partial t}\left(\rvar B\right)=\frac{\partial}{\partial\rvar}
  \left[\rvar\;\nu(B)\;\frac{\partial B}{\partial\rvar}\right]
  \label{eq:diff} 
\end{equation}
were investigated by numerical and analytical methods. Here, $B$ is the 
magnetic flux density, $\rvar$ is the radial coordinate, and $\nu$ is the 
turbulent magnetic diffusivity. For the dependence of the diffusivity on the 
magnetic field they considered a simple function satisfying the basic physical 
requirements:
\begin{equation} 
  \nu(B)=\frac{\nu_0}{1+|B/B_{\mbox{\scriptsize e}}|^{\alpha_{\nu}}} 
  \label{eq:nuexpr} . 
\end{equation}
Here $\alpha_\nu$ is a parameter quantifying the steepness of the diffusivity 
cutoff near $B_{\mbox{\scriptsize e}}$, the latter being the field strength 
where the diffusivity is reduced by 50\,\%. Physically, one expects 
$B_{\mbox{\scriptsize e}}\sim B_{\mbox{\scriptsize eq}}$, 
the kinetic energy density of photospheric turbulence. The solutions of 
Equation (\ref{eq:diff}) were found to be 
qualitatively similar to the case with constant diffusivity as long as the 
inhibition of turbulence was weak (i.e.\ for low values of $\alpha_\nu$ and/or 
$B_0/B_{\mbox{\scriptsize e}}$, $B_0$ being the central field strength of the 
initial field profile). For strong inhibition of turbulence however a 
new class of solutions sets in: a \it 
current sheet\/ \rm is spontaneously formed around the tube, moving inwards 
with a constant speed $w$, while the field strength inside the current 
sheet remains practically unchanged during the decay until the arrival of the 
current sheet. This leads to a parabolic decay of the cross 
section of the tube.

%We note that in a parallel work R\"udiger \& Kitchatinov (1996) solved the \it 
%2-dimensional\/ \rm nonlinear diffusion equation. 
%As far as it can be judged from their figures, their 
%solution falls in the diffusive regime, with a nearly linear decay law (in 
%contradiction to the observations, as shown in the analysis of the next 
%section). As however only results from a single run with one particular choice 
%of initial conditions are available, and the uppermost 1500 km of the 
%convective zone (where $B/B_{\mbox{\scriptsize {e}}}$ would be highest) is not 
%included in the 
%model volume, it is presently not possible to judge the general validity 
%of these findings.

%On the other hand, the models of Petrovay \& Moreno-Insertis (1996) 

An important and attractive feature of the erosion model is that the velocity 
$w$ of the current sheet becomes asymptotically independent of $\alpha_\nu$, 
i.e.\ for $\alpha_\nu\rightarrow\infty$ it goes to the finite limit
\begin{equation} 
   w\simeq  2^{-1/3}
   \frac{B_{\mbox{\scriptsize e}}}{B_0-B_{\mbox{\scriptsize e}}}
   \frac{\nu_0}{r_0} ,  \label{eq:wexpr} 
\end{equation}
where $r_0$ is the maximal radius of the spot.
Thus, the lifetime of the spot remains finite even for infinitely effective 
inhibition of the turbulence inside the tube. 

Using Equation (\ref{eq:wexpr}) and assuming that the unperturbed value of the 
diffusivity outside the tube is $\nu_0=1000$\,km$^2$/s, i.e.\ the granular 
value, $B_0=3000\,$G and $B_{\mbox{\scriptsize e}} =400\,$G, one finds that 
the Gnevyshev--Waldmeier rule is returned:
\begin{equation} 
   r_0/w \simeq 2^{1/3}
   \frac{B_0-B_{\mbox{\scriptsize e}}}{B_{\mbox{\scriptsize e}}}
   \frac{r_0^2}{\nu_0}=A_0/\overline D 
   \qquad \overline D\simeq 10   . \label{eq:Gnevipred}
\end{equation}
For the instantaneous decay rate $D=\dot A$ in turn we have the prediction
\begin{equation} 
  D\simeq\left(2^{2/3}\pi\nu_0
  \frac{B_{\mbox{\scriptsize e}}}{B_0-B_{\mbox{\scriptsize e}}}\right)
  \frac r{r_0} = C_D\frac r{r_0} \qquad C_D\simeq 22 \label{eq:Dlaw}  .
\end{equation} 

\begin{table}
\caption{Predictions of different sunspot decay models.}
\begin{tabular}{lcc}\hline
& Self-similar & Turbulent \\ 
& sunspot & diffusion \\ \hline
Decay law & $D=\,$const. & $D=\,$const \\
Lifetime & $T\propto A_0$ & $T\propto A_0$ \\
$B_0(t)$ & constant & gradual decrease \\
Spot boundary & sharp & diffuse \\
Reference & Gokhale \& & \citeN{Meyer+} \\
 & Zwaan (1972) & Krause \& R\"udiger (1975) \\ \hline
&&\\ 
&&\\ 
\hline
& Turbulent & Universal \\
& erosion & parabolic \\ \hline
Decay law & $D\propto r/r_0$ & $D\propto r$ \\
Lifetime & $T\propto A_0$ & $T\propto\sqrt{A_0}$ \\
$B_0(t)$ & constant & constant \\
Spot boundary & sharp & sharp \\
Reference & Petrovay \& & Mart\'{\i}nez Pillet \\
& Moreno-Insertis (1997) & {\it et al.} (1993)\\ 
\hline
\end{tabular}
\end{table}

The predictions of different published decay models are summarized in Table I.
The ``universal parabolic'' model in the table is identical to the turbulent 
erosion model except that the current sheet velocity $w$ is assumed to be a 
universal constant instead of depending on the maximal spot radius as 
in Equation (\ref{eq:wexpr}). This ``model'' has no theoretical foundation, and
it was only tentatively suggested by 
\citeN{VMP+:periph.decay} to forge a link between the lognormal distributions 
of decay rates and spot areas. 

\section{Data}
\subsection{Selection}
DPR data for the years 1977 and 78 (\citeNP{DPR}; \citeNP{DPR78}) were used, 
selecting only spots with  
maximal areas exceeding 10 MSH and measurements made within $\pm
72^{\circ}$ of the central meridian. In the following, this part of the 
solar surface will be referred to as ``the visible hemisphere''. For the 
present study only total
(umbra+penumbra) areas are used. Altogether, our data set consisted of 
3990 area measurements for 476 different spots.

The nominal error of measurements in the DPR is less than 1\,MSH, implying a
``worst case'' visibility function (\citeNP{Kopecky+:visibility})
$\mbox{sec}^2\,\lambda$ where $\lambda$ is the angular distance from the center
of the disk. Variable exposure times and seeing conditions, aggravated by the
geometric foreshortening correction factors for higher $\lambda$'s however lead 
to a larger random error in the area measurements. The Introduction of the DPR
(\citeNP{DPR}) quotes error values of typically 5\,\%, reaching 20\,\% for the
smallest spots in the sample. In spite of the great precision and care the DPR 
was compiled with, there is also a small chance
for errors in the spot identification, especially in  complicated, quickly
changing sunspot groups.
These errors are small compared to the random
physical scatter present in the data, and being random, they cannot contribute
to the physical correlations we find in Sections 4 and 5.

A \it systematic error\/ \rm is also involved owing to the
physical foreshortening (\citeNP{Archenhold}). While this error can be quite
significant for small spots close to the $\pm 72^\circ$ CM distance limits,
it will not exert a great influence on a sample as a whole unless the spots of
that sample have a strong tendency to be situated close to the limits. Such a
tendency is known to be present in at least one case. The first observation of
a spot on the ``visible hemisphere'' is considered as its \it birth\/ \rm if 
the spot was not seen at the
previous observing time (typically, the previous day). The \it death\/ \rm of a
spot is defined in an analogous manner. While the distribution of birth and
death sites has a well-known East--West asymmetry, peaking near the limbs, 
a straightforward estimate following the method of
\citeN{Kopecky+:visibility} shows that with the visibility function quoted
above the effect is small, and birth and death sites in our sample are
distributed nearly uniformly, i.e.\ the systematic error is small.

One should also be aware that some steps in the compilation of the data set
lead to \it selection effects. \rm An obvious effect is that 
the DPR database only includes 
spots for which at least 2 observations exist. Taking into account that
observations are typically separated by 1-day (and occasionally 2 to 4-day)
intervals, this implies that our sample is not complete below lifetimes 
of about 5 days. In what follows,
this effect will be referred to as the \it short lifetime selection effect. \rm

Another effect is the \it visibility reduction of lifetime\/ \rm
(\citeNP{Kopecky+:visibility}) which will cause a systematic error downwards in
determined spot lifetimes. The error is maximal for lifetimes of 8--10 days,
but with the visibility function at hand it altogether remains small.

\subsection{Definitions}
Let us assume that a given spot was observed at times $t_1,\dots,t_N$, the
corresponding area values being $A_1,\dots,A_N$. If $t_1$ and $t_N$ correspond 
to the actual birth and death of the spot, as defined in Section 3.1 above, 
then the \it lifetime\/ \rm of the spot is given by
\[
T\defdef t_N-t_1+1^{\mbox{\scri d}}
\]

For each measured area value the \it equivalent radius\/ \rm $r_i$ of the circle
having the same area is computed. In this paper,
equivalent radii will be expressed in MSHER units: a spot of radius 1 MSHER has
an area of 1 MSH, so that in these units $A_i=r_i^2$ (1 MSH$=3044\,$km${}^2$, 
1 MSHER$=984\,$km.) Time will be measured in days. The
spot is regarded to be in a decay phase at time $t_i$ if $r_{i-1}\ge 
r_i>r_{i+1}$ holds (or if $r_{N-1}\ge r_N$, in the case of the last
observation). The ``instantaneous'' decay rate of the spot is then given by
$D_i=2\pi r_i w_i$ where $w$ is the velocity  of the spot boundary, calculated
from
\begin{eqnarray}
 && w_i=
   w_i^{(0)}+2\frac{(w_i^{(1)}-w_i^{(0)})(t_i-t_{i-1})}{(t_{i+1}-t_{i-1})^2}
 \\
 && w_i^{(j)}=\frac{r_{i-1+j}-r_{i+j}}{t_{i+j}-t_{i-1+j}}
\end{eqnarray}
In the case of the last observation, $w$ is calculated for a time
$t_{N'}=(t_{N-1}+t_N)/2$ from
\[
w_{N'}=\frac{r_{N-1}-r_N}{t_N-t_{N-1}} .
\]
We note that the above formulae imply the assumption $w=\,$const.\ between
observations, i.e.\ a
parabolic decay law. In order to ensure that our results are not distorted by
such a tacit assumption, we repeated all statistical studies by using an
alternative definition for $D$, based on a linear decay law. The results of
Sections 4 and 5 were not significantly influenced by this choice. As those
results exclude the linear decay law, here we use the above formulae in all
calculations.

Similar care must be exercised with the definition of the maximal spot area 
$A_0$,
corresponding to an equivalent radius $r_0$, at time $t_0$. Many complex spots
show several local maxima (and consequently several decay phases) in their 
$A(t)$ curves, so the definition of a single maximum is not unique. For the 
results presented in this paper the maximum was defined as the \it absolute\/ 
\rm maximum \it preceding\/ \rm the given observation of the spot:
\[
A_{0,i}=\mbox{Max}\,\{ A_j\}_{j=1,\dots,i} \qquad A_0\equiv A_{0,N} . 
\label{eq:A0def}
\]
Again, using alternative definitions for the maximum, the results were not
significantly influenced (although using the preceding local maximum instead 
of the absolute maximum made most correlations somewhat worse). 

Another problem related to maximum determination is that owing to the rotation
of the Sun, in many cases we cannot follow the complete evolution of the spot.
In these cases we cannot be sure if the absolute maximum we determined is \it
really\/ \rm the absolute maximal area of the spot during its prior evolution.
In order to deal with this problem, the sample was divided into subsamples of
different ``reliability'' (e.g.\ spots born and died on the visible hemisphere
etc.). While using a more carefully selected sample does reduce the scatter in
statistical relationships, it may also reduce the significance of the result
owing to the lower number of data. Again, most of the studies presented in
Sections 4 and 5 were performed on different subsamples. The results presented
here use subsamples chosen to minimize the scatter while being still large 
enough to be statistically significant.

\subsection{Bias corrections}
The explicit selection effects mentioned in Section 3.1 above originate from 
conscious decisions during the compilation of the data set. Our sample is 
however also affected by \it implicit selection effects\/ \rm or \it biases.
\rm A bias here means that the distribution of some variable $y$ in our data 
set will differ from its actual distribution among sunspots because the
\it expected number of observations\/ \rm for one spot depends on $y$. Let $\hat
P(y;\vc x)$ denote the observed distribution of $y$ in some subset of our data
homogeneous in the variable(s) symbolically denoted here by $\vc x$. If the
real distribution (i.e.\ the ``spot distribution'' as opposed to the ``data
distribution'') is denoted by $P(y;\vc x)$ and the expected number of data for 
one spot by $p(y;\vc x)$ then clearly
\[ P(y;\vc x)=F(y;\vc x)\hat P(y;\vc x) \]
where $F\propto 1/p$, with some arbitrary normalization. It is convenient to
choose the normalization so that in our discrete sample of $N$ data,
$\sum_{i=1}^N F_i=N$. Formulae for $p$ viz.\ for the correction factor $F$ for 
cases of practical importance will be given below.

Often we are only interested in the mean and the standard deviation of $P$.
These and their r.m.s.\ errors are given by
\begin{eqnarray}
&& \ov y=\langle yP\rangle=\frac 1N\sum_{i=1}^Ny_iF_i \label{eq:binmean}\\
&& \sigma^2=\langle (y-\ov y)^2P\rangle =\frac 1{(N-1)}\sum_{i=1}^N(y_i-\ov y)^2F_i \\
&& \sigma_{\ov y}^2=\sigma^2\sum_{i=1}^N\left(\pdv{\ov y}{y_i}\right)^2=
   \frac{\sigma^2}{N^2}\sum_{i=1}^N\left(F_i+y_i\pdv{F_i}{y_i}\right)^2
   \label{eq:binmeansigma} \\
&& {\sigma_{\sigma}^2=\sigma^2\sum_{i=1}^N\left(\pdv{\sigma}{y_i}\right)^2=}
 \nonumber \\
&& \quad = 
   \frac 1{4(N-1)^2}\sum_{i=1}^N\left[2(y_i-\ov y)F_i+(y_i-\ov y)^2\pdv{F_i}{y_i}
   \right]^2 
   \label{eq:binsigsig}
\end{eqnarray}

If one wishes to make a least-squares fit $y=y_0(\vc x)$ to the data, the
simplest way to do this for a biased sample is to bin the data in $\vc x$ 
using
bins small enough that $y_0$ may be securely assumed to be nearly constant
within each bin. Then $\ov y$ is calculated for each bin from Equation
(\ref{eq:binmean}), and the fit is performed on these bin averages, weighting
each value by the inverse of its $\sigma_{\ov y}^2$, as given by Equation
(\ref{eq:binmeansigma}). 

\subsubsection{$y=f(T)$: Gnevyshev--Ringnes correction}
The lifetime of a spot can obviously only be determined if both its birth and
death occur in the visible hemisphere, so only such spots will enter our data
set. The probability $p$ of this being so (i.e.\ the expected number of data
for one spot) however depends on $T$
according to a formula first given by \citeN{Gnevyshev} and later corrected
by \citeN{Ringnes:correction}. For our purposes this can be written as
\[
p=\left\{ \begin{array}{ll}
         0 & \mbox{if}\quad |\alpha |\ge 2\theta_0 \\
         &\\
	 \dfrac{2\theta_0-|\alpha |}{360^{\circ}}\quad &  
	    \mbox{if}\quad |\alpha | < 2\theta_0 
	  \end{array} \right.
    \label{eq:GRcorr}
\]
\[ \alpha\defdef \left[ 
   180^{\circ}-(\Omega T+180^{\circ})\,\mbox{mod}\,360^{\circ} \right] 
\]
Here, $\theta_0=72^\circ$ is the CM distance limit and  
$\Omega=360^\circ /P_{\mbox{\scri syn}}$, $P_{\mbox{\scri syn}}$ being the
synodic rotation period of the given spot. For simplicity, here we use the
value $P_{\mbox{\scri syn}}=28^{\mbox{\scri d}}$ for all spots. (In
1977/78 sunspots were situated at high latitudes.) 

The correction factor $F$ is the normalized inverse of Equation 
(\ref{eq:GRcorr}), independently of the form of the function $f(T)$. Its
derivative $F'(y)$, entering Equations (\ref{eq:binmeansigma}) and 
(\ref{eq:binsigsig}), will obviously depend on that form, though. 
In Section 5 below we will deal with the distribution of $y=\log T$: in that
case $F'(\log T)=\ln 10\, T F'(T)$.

\subsubsection{$y=f[D;g(A, A_0)]$: correction for the decay law}
Let us first restrict our interest to the case $g(A, A_0)=A$. In this case our
study of the distribution of $y$ is limited to spots in the area interval
$[A,A+dA]$. A spot with a decay rate $D$ will spend a time $dA/D$ in this
interval, so if our observations are made with $1^{\mbox{\scri d}}$ intervals,
the expected number of observations for one spot in that interval (in our units
of measurement) is clearly
$p=dA/D$. In the more general case
\[ p=\dv Ag\frac{dg}D   . \]
It is apparent that $p$ (and $F$) here depend on the form of the $g(A, A_0)$
function. This is in contrast to the Gnevyshev--Ringnes correction where the
observation probability of a spot of lifetime $T$ was independent of any other
spot parameter than $T$. For the two cases investigated in Section 4 below the 
correction factor takes the form
\[
F=\left\{\begin{array}{ll} F_0\dfrac D{A_0g} \quad &\mbox{if}\quad 
                         g=(A/A_0)^{1/2} \\
			 &\\
			 F_0\dfrac D{g} \quad &\mbox{if}\quad 
			 g=A^{1/2} \end{array}
  \right. 
\]
$F_0$ being a normalization factor.
$F$ is again independent of the form of $f(D)$ while its derivative for $y=\log
D$ is $F'=\ln 10 F$.

\subsubsection{Limitations of the bias correction}
Assume that all our measured $y_i$ data fall within an interval 
$J$. If $p\gg 1/N$ in the neighbouring intervals of similar size
then we can be reasonably sure that the lack of data points outside the above
range is not an artefact caused by the bias. (This is the case with our
studies in Section 4.) If however the range of strong bias $p\la 1/N$ overlaps
$J$ or is immediately adjacent to it, we should suspect that the 
lack of data in the strong bias range is an artefact. In this case, no proper 
correction is possible. The values derived  for the 
mean and standard deviation, however, may still be acceptable [albeit with the 
large error bar given by Equation (\ref{eq:binmeansigma})] if data points are 
present both above and below the strong bias range, thereby offering the 
possibility of an interpolation. This is the case we will encounter in 
Section~5.

Beside the implicit selection effects dealt with above, there is a variety of
subtle biases affecting certain distributions, the corrections of which is not
always straightforward or worth the effort. We will meet two such examples in
the following section. Unfortunately, neither these more involved selection
effects nor 
others discussed in Sections 3.1 and 3.3 were properly taken into account in 
much of the earlier work on the subject. 

\section{Decay law}
\subsection{$D$ vs.\ $r/r_0$}
\begin{figure}
%\vspace{1 cm}  % Amount of vertical space needed
\centerline{\psfig{figure=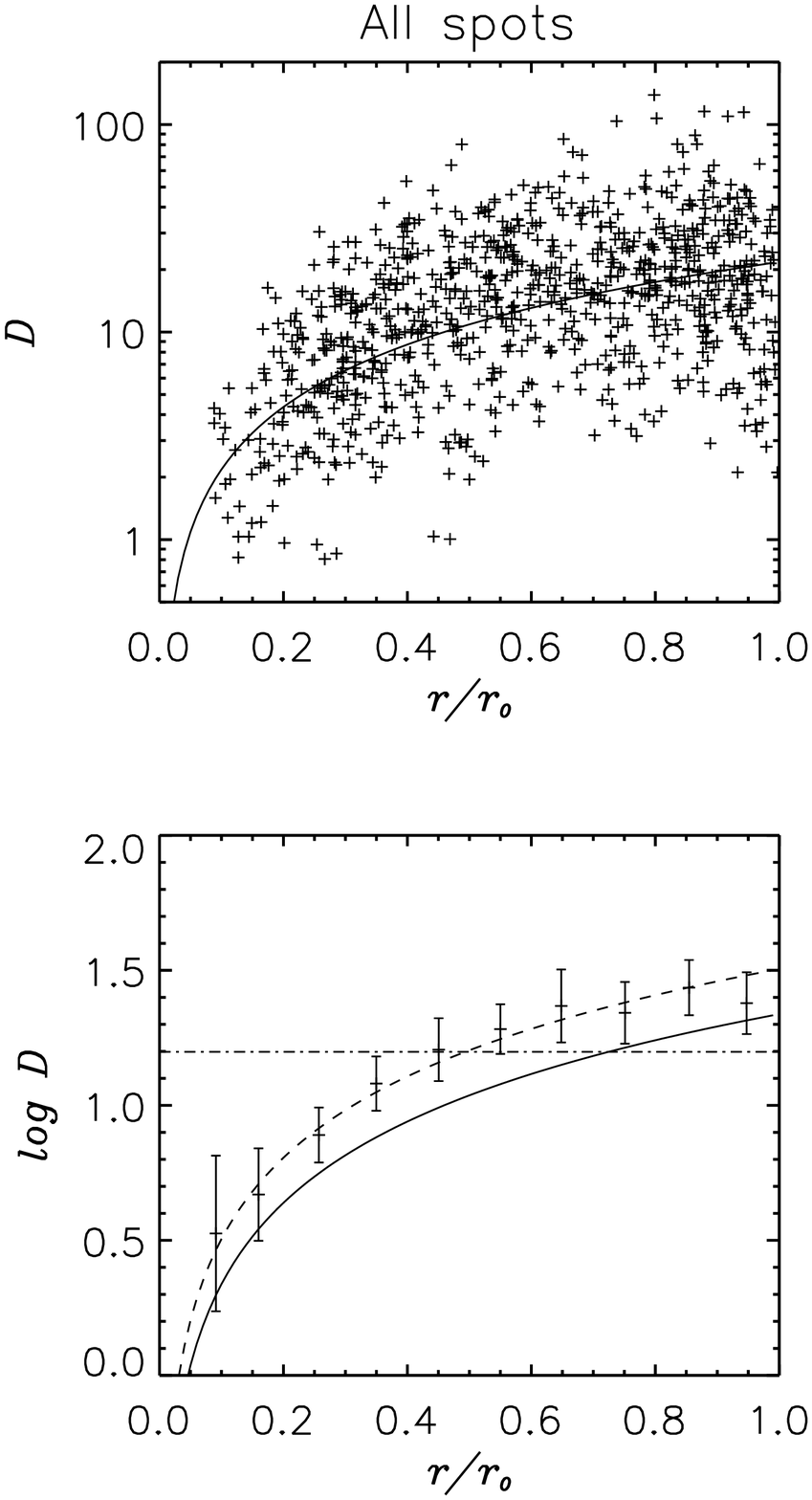,width=7.0 cm}\hskip -0.6 cm
            \psfig{figure=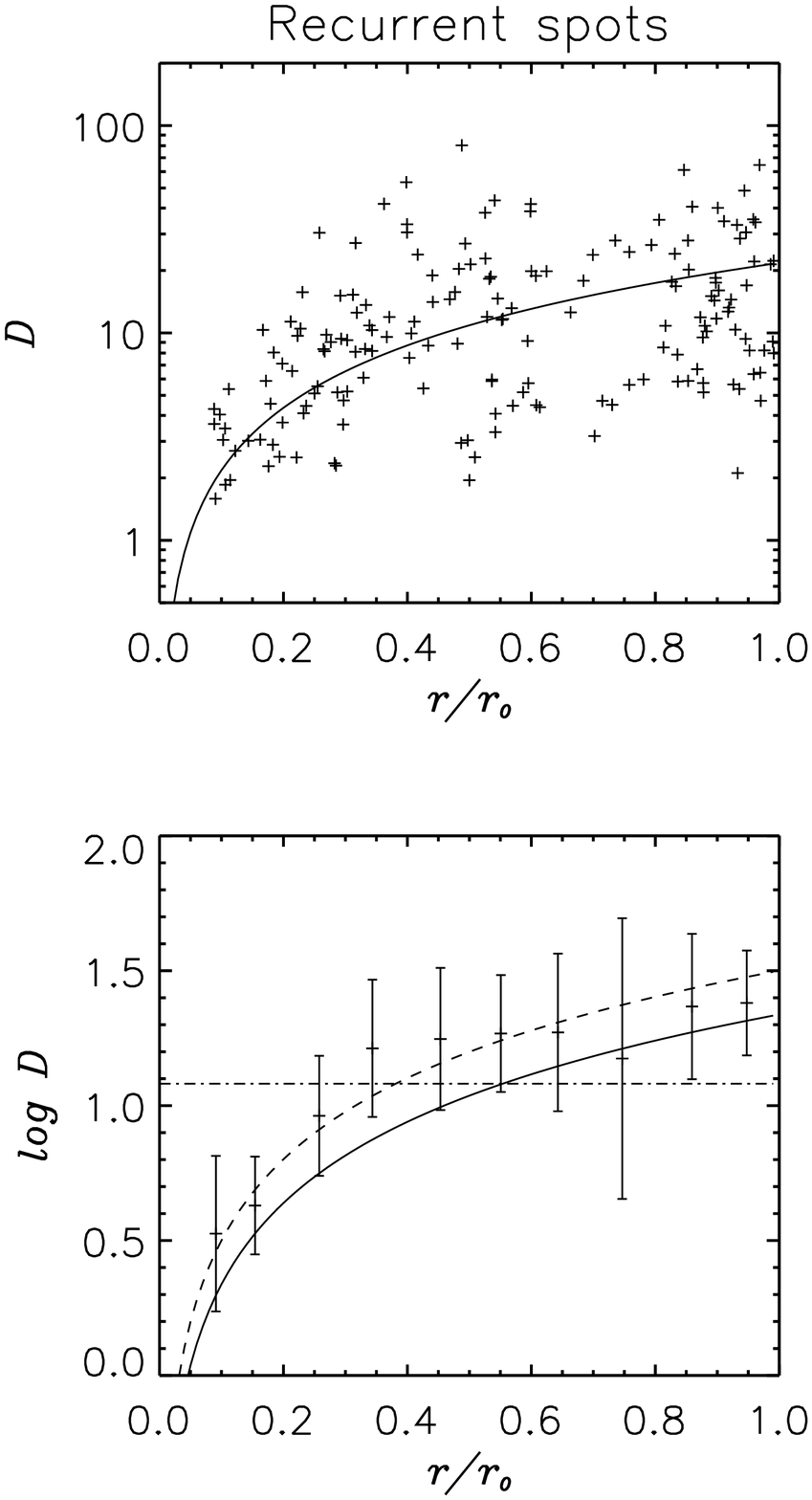,width=7.0 cm}}
\caption{Decay rates plotted against 
relative radius for the original (top) and binned (bottom) data.  Data for
spots which died on the visible hemisphere are shown. Solid line: 
Equation~(8). Dashed line: linear least-squares fit.
(Here and in all other figures, $2\sigma$ error bars for the mean are shown.)}
\label{fig:Dlaw}
\end{figure}

Guided by the prediction of the turbulent erosion model, Equation
(\ref{eq:Dlaw}), in Figure \ref{fig:Dlaw} we plot the instantaneous decay rates
$D$ against the relative radius $r/r_0$ [in the sense of $r_i/r_{0,i}$, cf.\
Equation(\ref{eq:A0def})]. In this plot we show measurements for
those spots which died on the visible hemisphere (888 data points, 170 of these
for recurrent spots). The results for the complete sample are however very
similar. As sunspot decay rates are 
lognormally distributed (\citeNP{VMP+:periph.decay}; see also the
following paper in this series), for the application of Gaussian-based 
statistical methods (least squares etc.) $D$ should be plotted on a 
logarithmic scale. 

It is apparent that the formula 
(\ref{eq:Dlaw}) is indeed a reasonable representation of the data. Binning the 
data in $r/r_0$ and correcting the bias described in Section 3.3.2 we find 
that a constant $D$ (i.e.\ a linear 
decay law) can be excluded at a very high confidence level 
($Q=\Gamma[(N-1)/2,\chi^2/2]<10^{-5}$). In contrast, a linear fit of 
the form $D=C_D\, r/r_0$ (i.e.\ a parabolic decay law, the best fit being 
$C_D=32.0\pm 0.26$) yields an excellent agreement ($Q=0.68$). 
The difference in the coefficient compared to Equation (\ref{eq:Dlaw}) can be 
easily accommodated taking into account the uncertainty 
of the value of $\nu_0$, the strong height-dependence of 
$B_{\mbox{\scriptsize e}}$, and the fact 
that our formula (\ref{eq:Dlaw}) was based on the infinite $\alpha_\nu$ limit 
(\ref{eq:wexpr}), so the actual value of $w$ may be somewhat higher.

\begin{figure}
%\vspace{1 cm}  % Amount of vertical space needed
\centerline{\psfig{figure=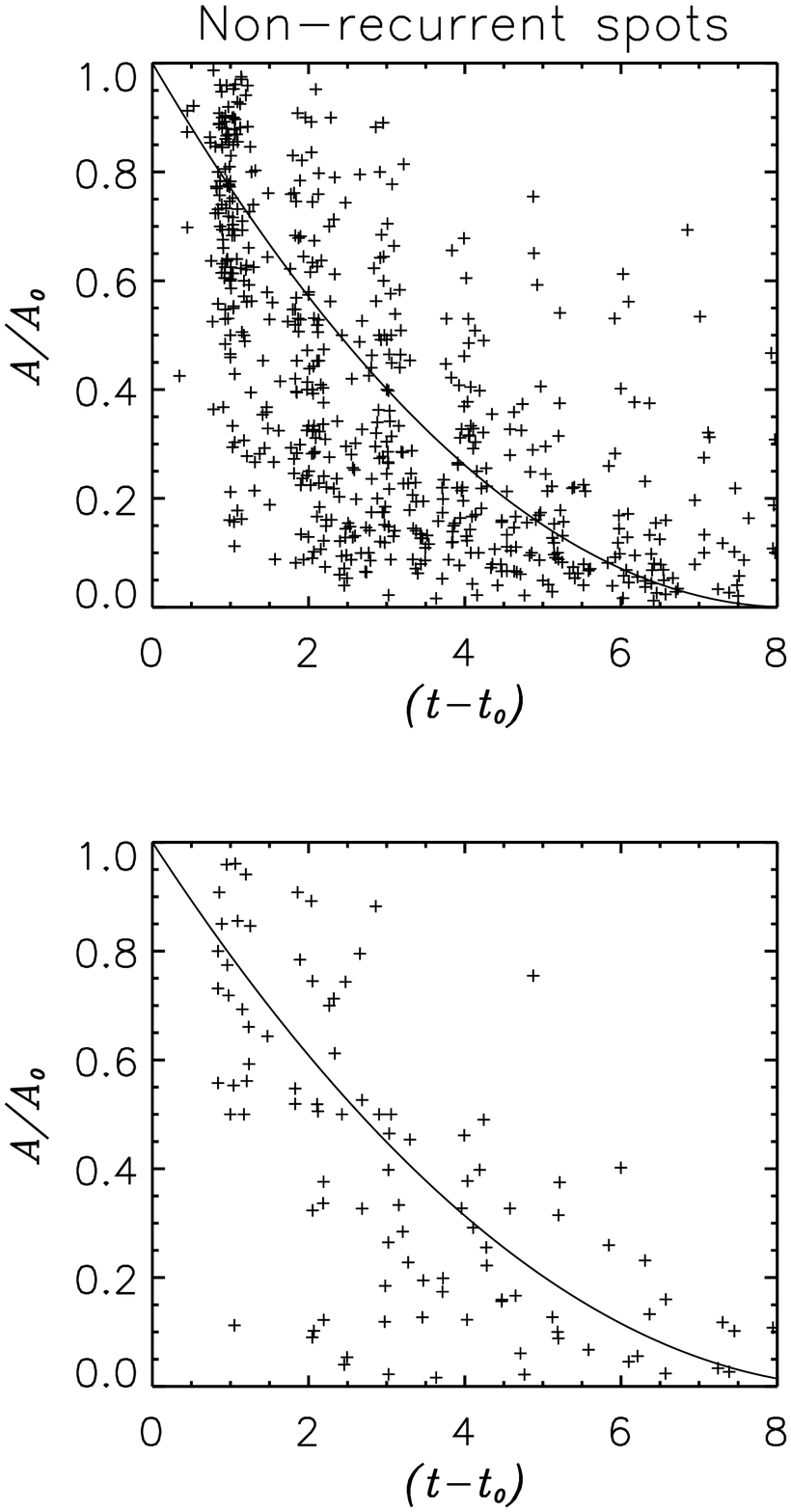,width=7.0 cm}\hskip -0.6 cm
            \psfig{figure=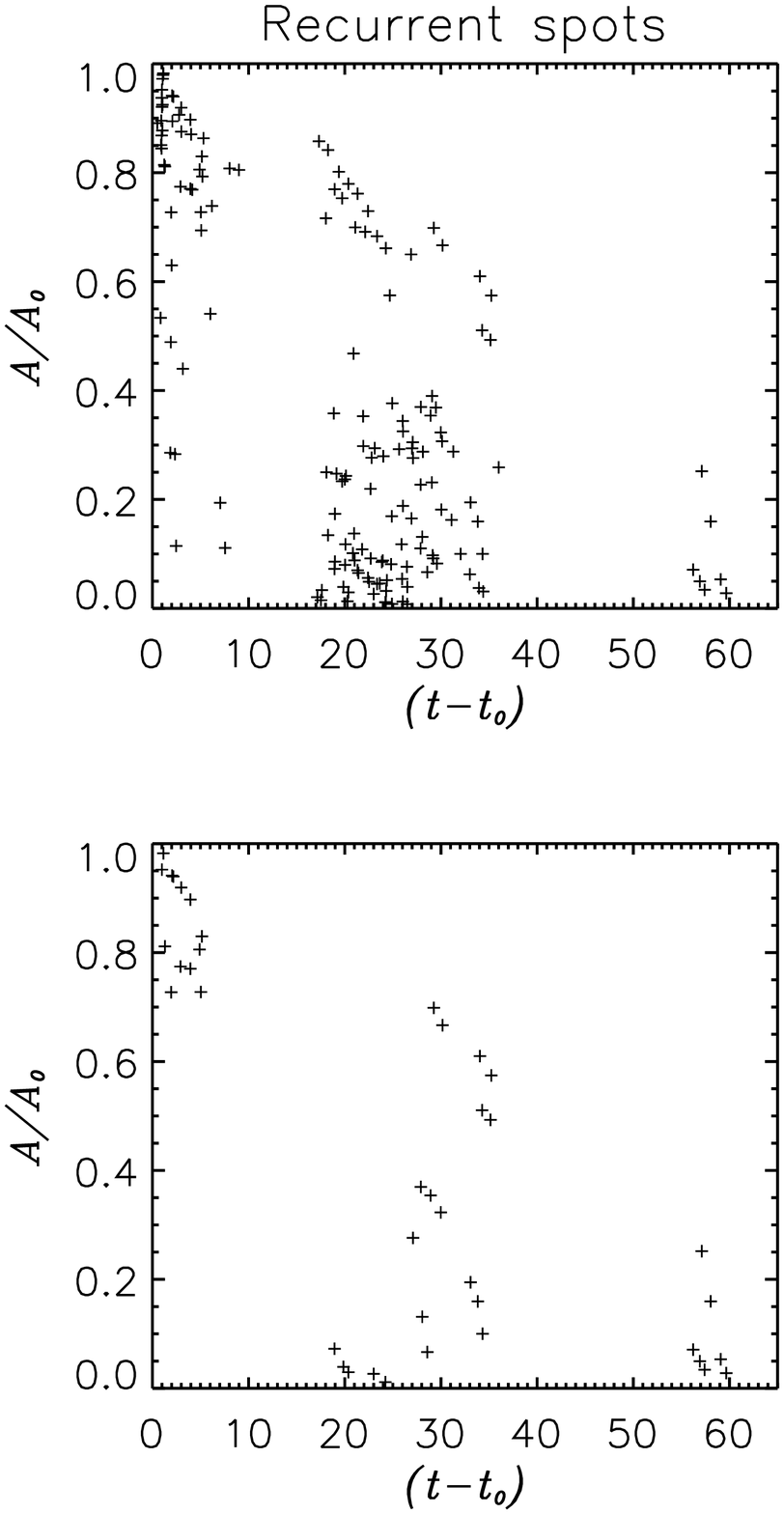,width=7.0 cm}}
\caption{Relative area vs.\ time plots. \it Top: \/\rm all spots; \it bottom: 
\/\rm spots with maximal areas in a restricted range (90--110 for 
non-recurrents, 250--350 for recurrents). Solid line: Equation (3) with $A_0=90$
and $w$ from Eqs.\ (6) and (8)}
\label{fig:Dcurve}
\end{figure}

The question arises, how is it possible that this obvious nonlinearity of the
decay had not been noticed earlier? While our use of the DPR with its high
precision individual sunspot area data may be a factor in a more
conclusive result, the study of \citeN{FMI+MVA} shows that with proper methods 
of analysis a nonlinearity can also be detected in the GPR data. We should
therefore suspect that the choice of the technique of analysis is a crucial
factor in this respect. Sunspot decay curves are usually charactarized by bad 
sampling intervals: for
short-lived spots one often only has 2--3 measurents in the decay phase, while
for recurrent spots a large number of measurements are grouped together in the
vicinity of 2--3 points of the decay curve. The true endpoint of the decay
curve is thus ill determined and the method of producing ``typical'' decay
curves by time-shifting individual curves to a common endpoint (e.g.\
\citeNP{Ringnes:shortspots}, \citeyear{Ringnes:longspots}) implies rather large
errors. These errors are further aggravated by the fact that the 
lifetime--maximal
area correlation (cf.\ Section~5) involves a large scatter, thus spots with
greatly different $r_0$ values will enter the sample, leading to different decay 
rates at the same time. The resulting large scatter in $D$ will tend to smear 
out any nonlinearity. This is borne out in Figure~\ref{fig:Dcurve} where the 
parabolic decay curve is directly seen in the subsample with $A_0$ in a narrow 
range (bottom left panel), while the picture is much more confuse if no such 
selection is performed (top left panel). (Note that some non-linearity is still 
visible even in this latter panel. This may be due to our time-shifting the
curves to a common maximum instead of a common endpoint, as usual.) The 
selection of an
$A_0$-limited subsample however greatly reduces the sample size making it
difficult to  establish the nonlinearity at a high confidence level for all but
the most frequent $A_0$ values. Thus, the $D$--$r/r_0$ plot
(Figure~\ref{fig:Dlaw}) is much more useful for the study of the decay law.

The study of \citeN{Bumba:decay} resulted in a nonlinear decay law for the
majority of non-recurrent spots while a linear decay law was found to describe
the recurrent spots. To check the reality of such a distinction, in the
right-hand panels of Figure~\ref{fig:Dlaw} we present the $D$--$r/r_0$ plot for
recurrent spots only. Essentially the same results are found as for the
complete sample. This is so despite the fact that this plot is distorted by an 
uncorrected
bias: the apparent local minimum near $r/r_0\simeq 0.7$ is caused by the fact
that between the first and second disk passages, higher-$D$ spots shrink more
---so the corresponding gap in $r/r_0$ will be wider in the upper part of the
diagram. Bumba's finding could be due to the fact that the $A_0$ values of 
recurrent spots are scattered within very wide limits. This, together with the
bad sampling intervals, leads to a very confuse area--time plot (righ hand
panels in Figure~\ref{fig:Dcurve}), for which a linear fit is as good as
anything else.

\begin{figure}[htb]
%\vspace{1 cm}  % Amount of vertical space needed
\centerline{\psfig{figure=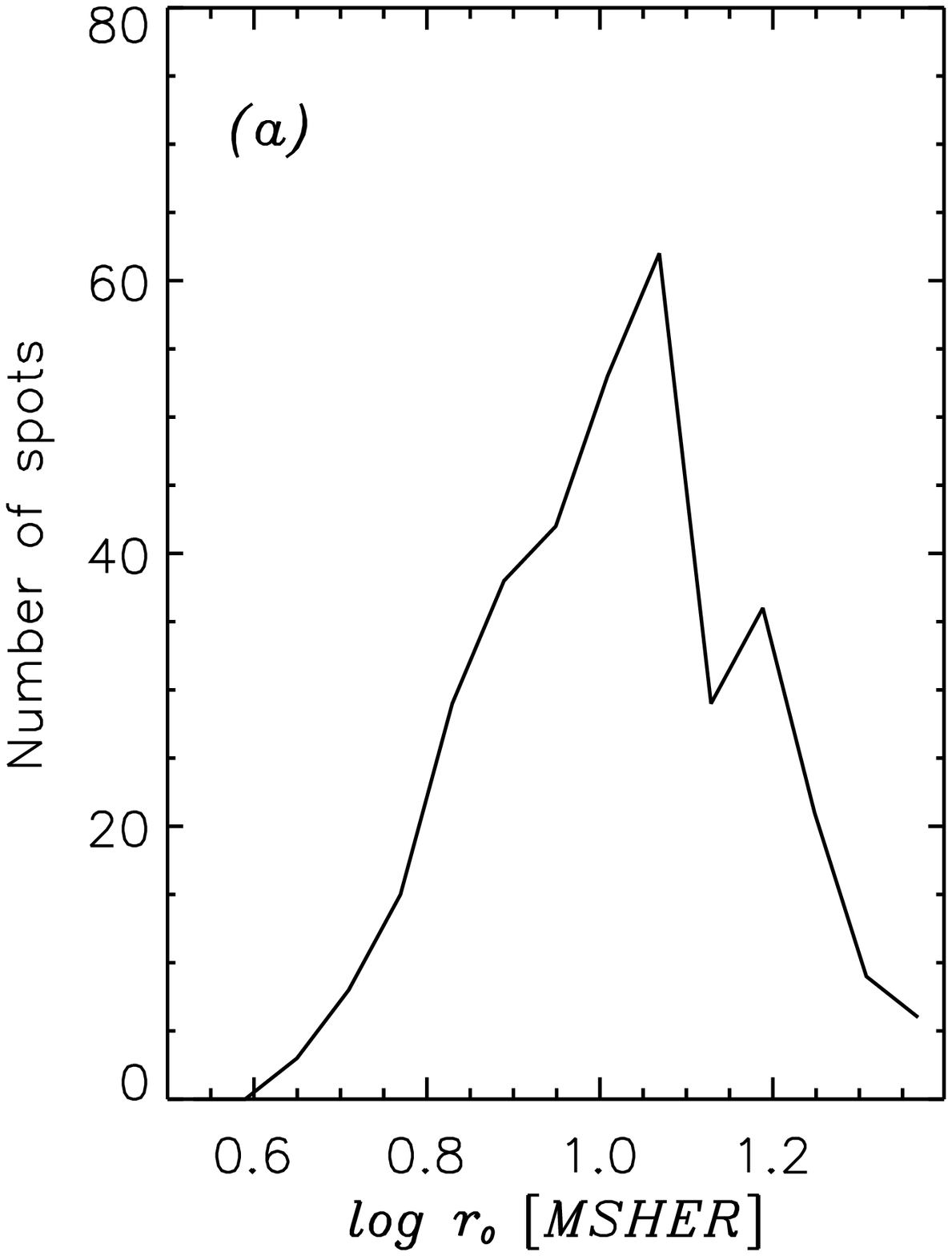,width=7.0 cm,height=7 cm}\hskip -0.6 cm
            \psfig{figure=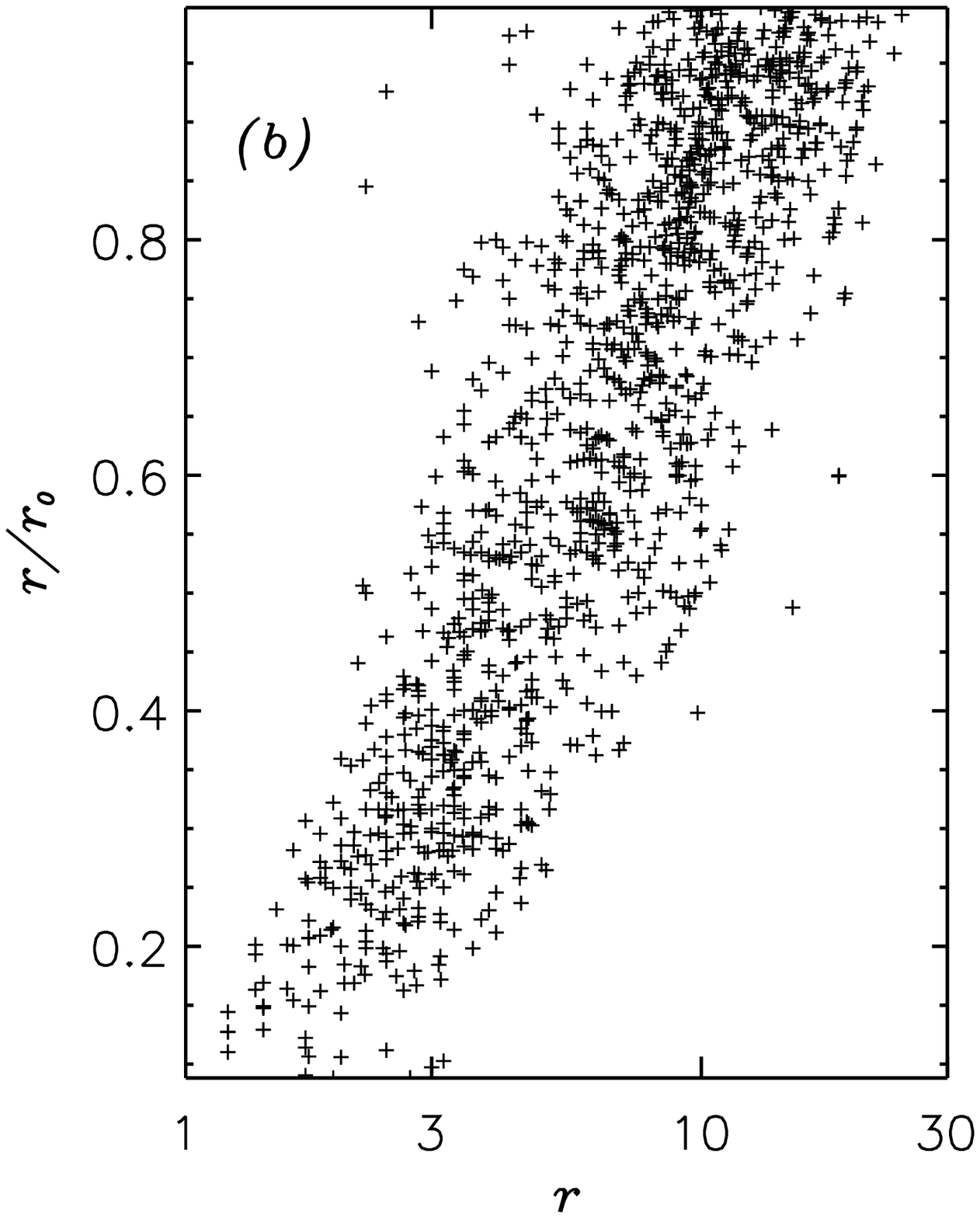,width=7.0 cm,height=7 cm}}
\caption{Histogram of maximal radii (a) and $r$ vs.\ $r/r_0$ plot (b) 
for all spots in our data set.}
\label{fig:rr0}
\end{figure}

\subsection{$D$ vs.\ $r$}
Having excluded the linear decay law and shown that $D\propto r/r_0$ is
compatible with the data, we still need to check the validity of the prediction
of the fourth model in Table I, i.e.\ the``universal parabolic'' decay law
$D\propto r$. This is necessary because in our data there is a good, nearly
linear correlation between $r$ and $r/r_0$, caused by the rather sharp peak in
the distribution of $r_0$ values, as illustrated in Figure~\ref{fig:rr0}.
(Note that this histogram is contaminated by several selection effects, but
this is irrelevant for our present purpose.) The suspicion arises that the
correlation between $D$ and $r/r_0$ found above might be just a reflection of a 
more basic relationship between $D$ and $r$.

\begin{figure}[htb]
%\vspace{1 cm}  % Amount of vertical space needed
\centerline{\psfig{figure=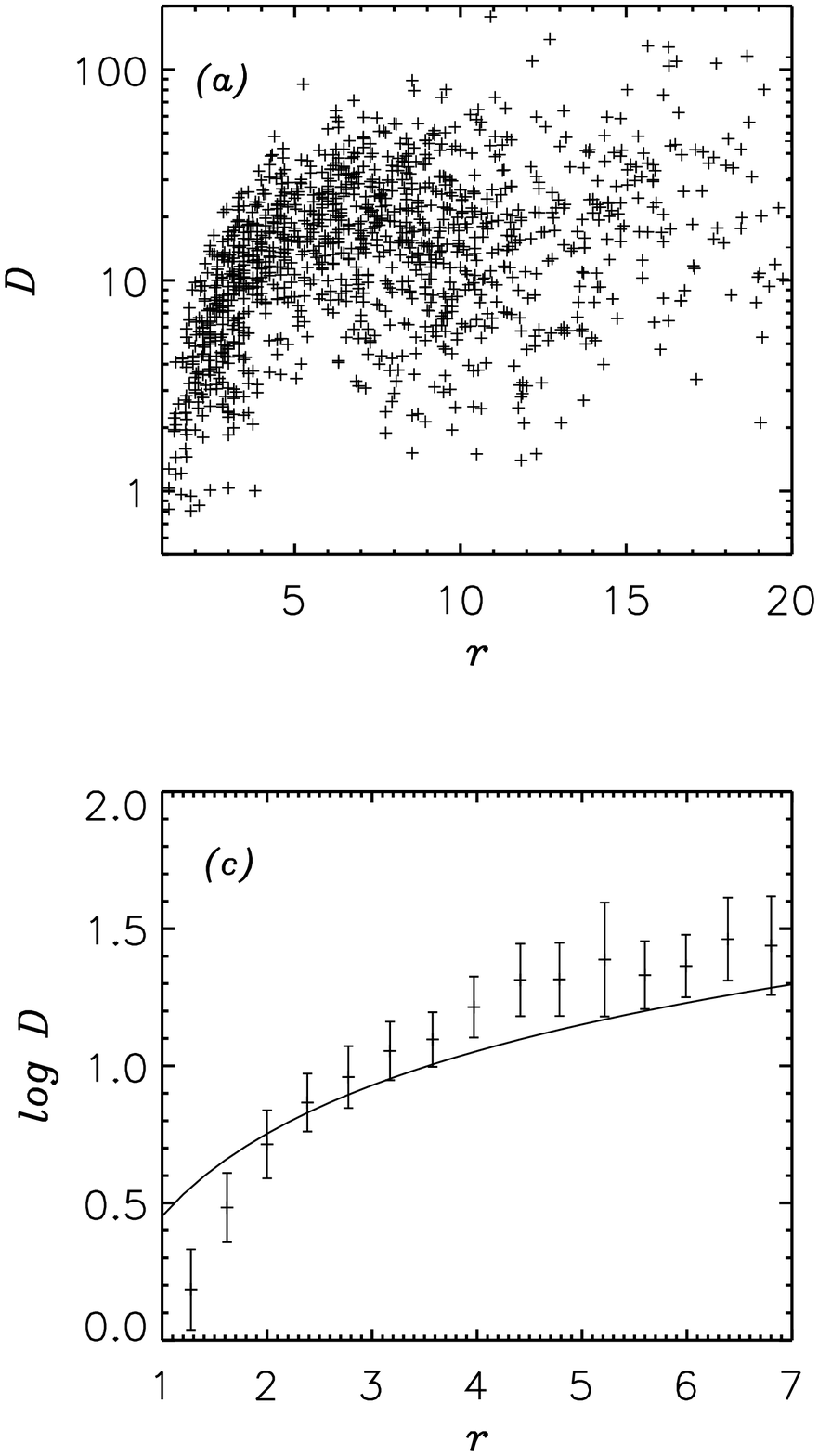,width=7.0 cm}\hskip -0.6 cm
            \psfig{figure=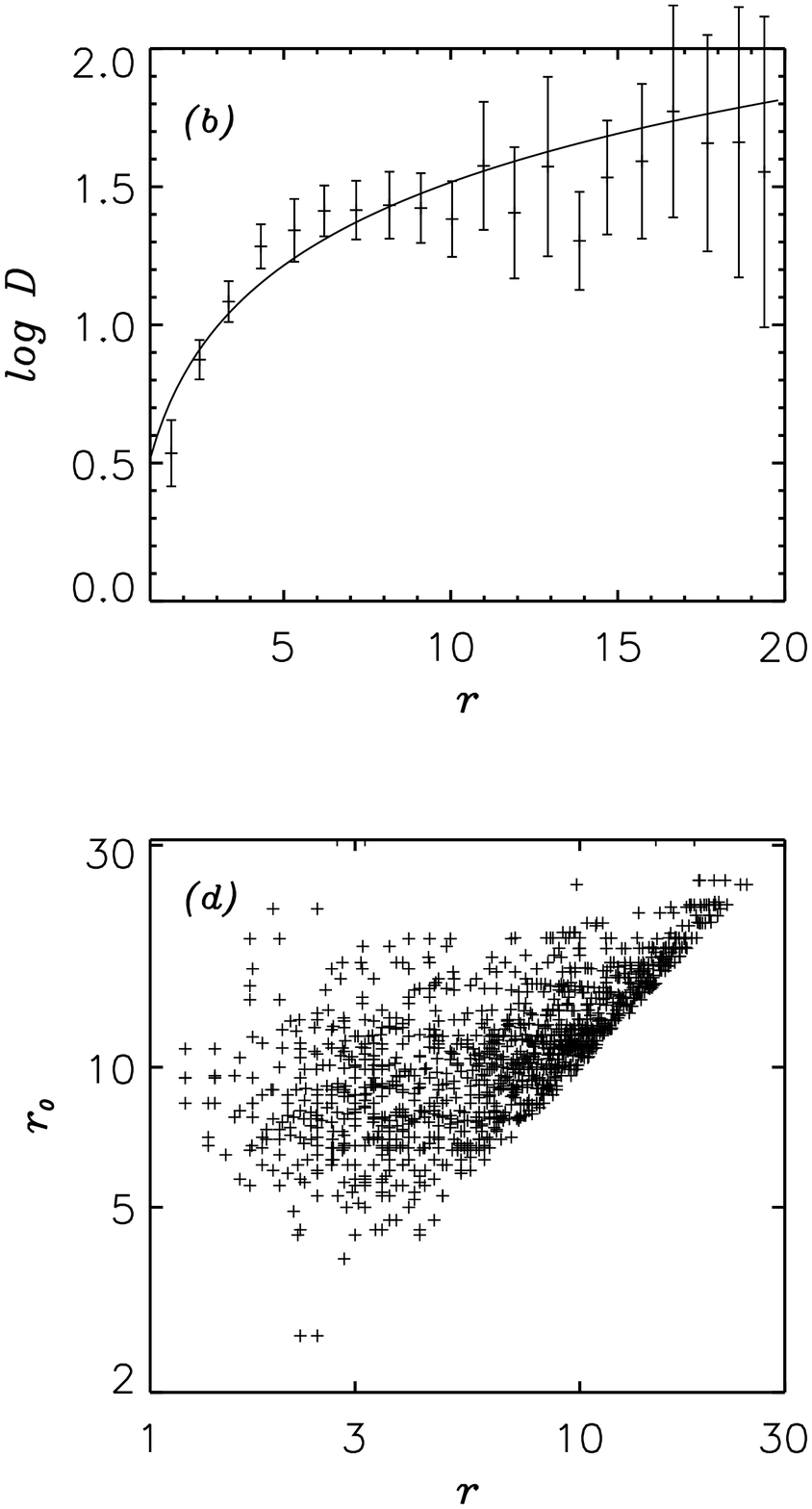,width=7.0 cm}}
\caption{$D$--$r$ plot for the original (a) and binned (b) data.  (c): low $r$
part with higher resolution; (d): $r$ vs.\ $r_0$ plot. All
decay rate measurements are shown. Solid line: linear least-squares fit.}
\label{fig:universal}
\end{figure}

The distribution of the data points from the full sample (1234 points) on the 
$r$--$D$ 
plane is shown in Figure~\ref{fig:universal}. It turns out that a fit of the
form $D\propto r$, corresponding to the universal parabolic model, is
incompatible with the data ($Q<10^{-5}$), as is a constant decay rate. 

The actual distribution of the data finds its explanation in 
Figure~\ref{fig:universal}d. For $r\ga 7$, there is a nearly one-to-one
correspondence between $r$ and $r_0$, i.e.\ most of these measurements refer to
spots close to their maximal radii. This is a consequence of the rather fast
decrease of the $r_0$ distribution curve, Figure \ref{fig:rr0}a, in this 
regime.
According to Equation (\ref{eq:Dlaw}), the decay rate is then constant: indeed,
for $r\ga 7$ the $D$--$r$ plot is compatible with $D=\,$constant. The sharp peak 
in
the histogram of $r_0$ near 10 also implies that for lower values of $r$ most of
the measurements are for spots with $r_0\simeq 10=\,$const. Thus, as it follows
from Equation (\ref{eq:Dlaw}), in the range $r=3$--$7$ the $D$--$r$ 
distribution is reasonably well fit by the linear relation $D\propto r$.

In the range $r\la 3$ the distribution again seems to depart from the linear 
law, this time being steeper. This is the consequence of a subtle
bias: the 1-day sampling interval is relatively ``coarser'' for small (and thus
short-lived, cf.\ Section~5) spots which are thus less likely to be ``caught''
in the latest stages of their decay. Consequently, the mean value of $r_0$ 
begins to increase again with decreasing $r$ in this regime, Figure
\ref{fig:universal}d.

The fact that the $D$--$r$ relation for $r>7$ is compatible with a constant
decay rate implies that we have no significant evidence for the existence of a
supergranular area ``quantum'' where spots are supposed to be more stable, 
decaying
slower. An independent test of this was also performed by taking the power
spectrum of the histogram of $r$ (not shown here): again, no significant peaks
were found. The area histogram published by \citeN{Howard:ARAA} leads to 
similar
conclusions. While it cannot be excluded that using a much larger database, such
an effect may be discovered in the future (note the slight local minimum in the
$D$--$r$ plot near $r\simeq 14$), none of the papers that claimed to have
found such a quantization was in fact based on a database significantly more
extensive than ours in the given area range ($\ga 80\,$MSH). Some of these works
used data for ``regular'' spots only, the selection criteria however were
rather subjective. Facing this state of the matter, we are forced to conclude 
that no compelling evidence
exists for the supposed phenomenon of quantized sunspot areas. 

\begin{figure}[htb]
%\vspace{1 cm}  % Amount of vertical space needed
\centerline{\psfig{figure=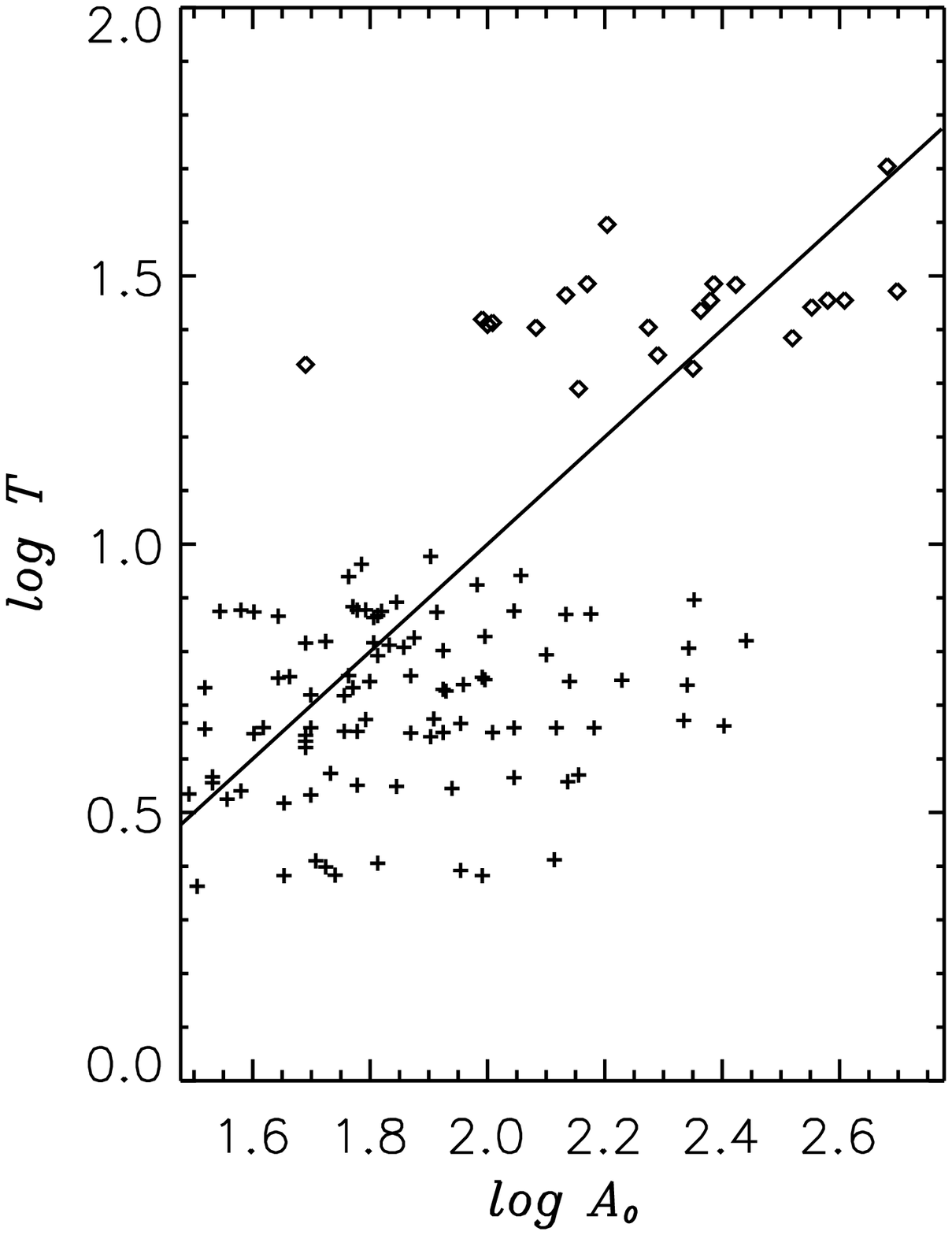,width=7.0 cm,height=7 cm}\hskip -0.6 cm
            \psfig{figure=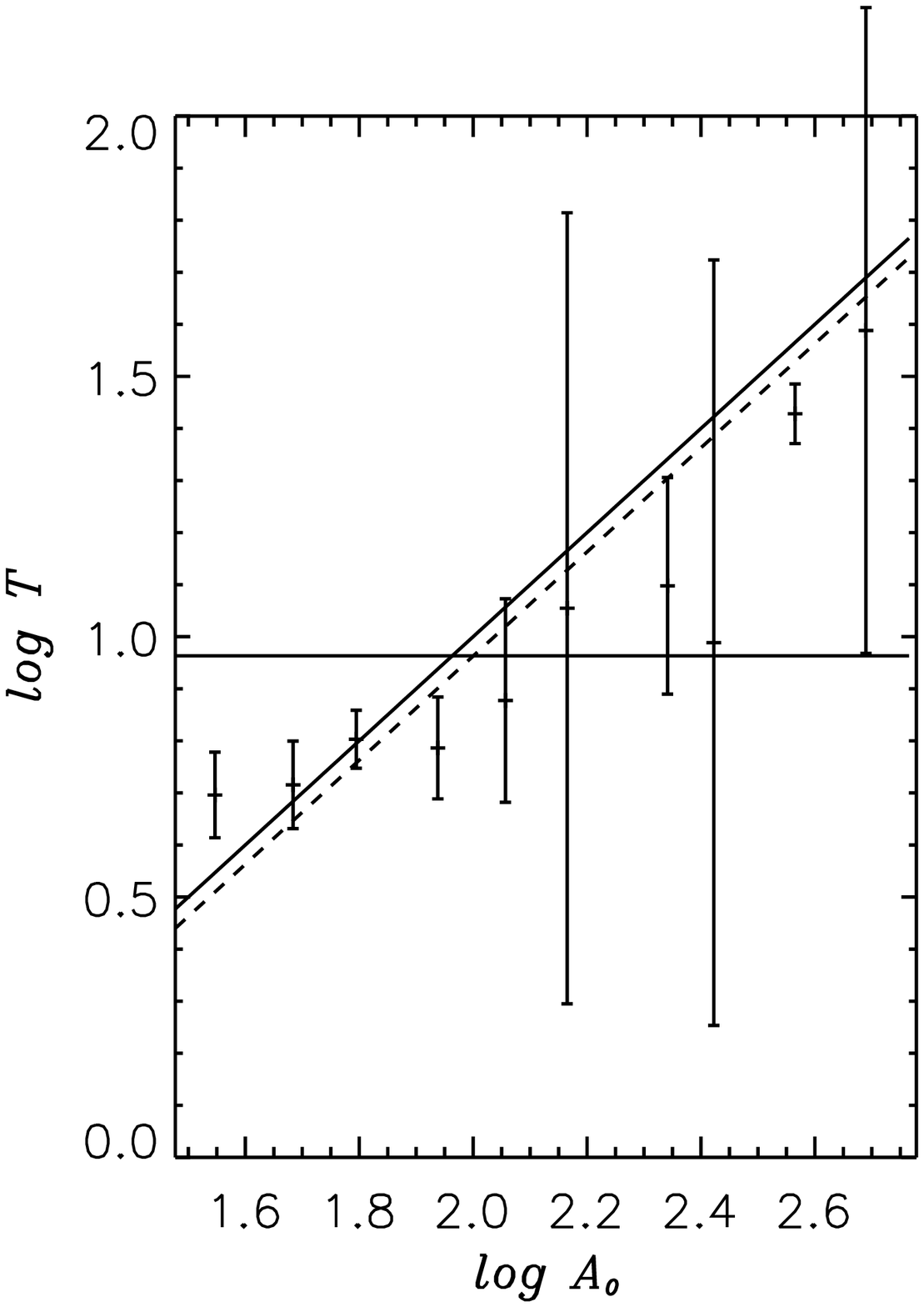,width=7.0 cm,height=7 cm}}
            \caption{Area--lifetime relation for all spots with determined lifetimes in the
1977/78 DPR data. $+$: non-recurrent spots; $\Diamond$: recurrent spots.
Solid line: Gnevyshev--Waldmeier relation $A_0=10\,T$; dashed
line: linear best fit.}
\label{fig:G-W}
\end{figure}

\section{Area-lifetime relation}
There are 128 spots in our sample which were born on the visible hemisphere and
also died there, so that their lifetime could be determined. These lifetimes
are plotted against the maximal radii in Figure \ref{fig:G-W}. The empty
horizontal band corresponds to spots living for about half a solar rotation:
obviously, no lifetime could be determined for any of these. After binning
and applying a Gnevyshev--Ringnes correction (Section 3.3.1), a linear fit
$A_0=\DGW T$ yields $\DGW =10.89\pm 0.18$, in excellent agreement with Equation
(\ref{eq:GW}). The validity of the Gnevyshev--Waldmeier rule is thus also 
proven for individual sunspots. 

Note that from Equation (\ref{eq:Dlaw}) with the value $C_D=32$ derived in the
previous section a mean decay rate of $\ov D =16$ would follow. This does not
contradict to the value of $\DGW$ derived here, as here we are dealing with the
ratio of the maximal area to the \it full\/ \rm lifetime of the spot which is
typically larger than the decay time by about 40\,\% (\citeNP{Vitinsky+:book}).

The data obviously show a rather large scatter around the linear law, and the
presence of exclusion zones corresponding to lifetimes around half-integer
rotations makes it even more difficult to discern the correlation, particularly 
for samples limited in $T$ ---e.g.\ \citeN{Vitinsky:spotcorr}. However, a
mean lifetime independent of maximal area, $T=\,$constant, is definitely 
excluded by the data ($Q<10^{-5}$, as opposed to $Q=0.18$ for the linear best 
fit).

Some explanation may be in order as to our choice of $A_0$ as independent
variable for the fit. Using $T$ as independent variable, as previous
authors did, would clearly make it possible to avoid the use of a
Gnevyshev--Ringnes correction, and it would not be sensitive to the short
lifetime selection effect (Section 3.1): this latter effect leads to the slight
upwards deviation of the leftmost binned data point in Figure \ref{fig:G-W}, 
and it is the reason why the data were cut down at
$A_0=30$ in the figure. These advantages of a $T$-dependent fit are however
offset  by the less uniform coverage of the abscissa in that case (the low
number of recurrent spots would lend a very small weight to those points in the
fit) as well as by the possible strong influence of the visibility reduction of
lifetime (Section 3.1) for some $T$ values (mainly for those just below the 
gap). This
influence may explain the suggestion by \citeN{Alexander} and 
\citeN{Ringnes:shortspots} 
that groups with a lifetime of 8 days significantly depart from the linear
law. 
In our material we also find somewhat similar
deviations near the exclusion zone, but the corresponding large error bars
indicate that these are not significant. 

\section{Conclusion}
In our statistical study of the decay of individual sunspots based on DPR data
for the years 1977--78 we found that the instantaneous area decay rate is
related to the spot radius $r$ and the maximal radius $r_0$ as
\[ D=C_D \,r/r_0 \qquad C_D=32.0\pm 0.26 . \]
This implies that sunspots on the mean follow a parabolic decay law; the
traditional linear decay law is clearly excluded by the data. A ``universal''
decay law of the form $D\propto r$ is also excluded. The validity of the 
Gnevyshev--Waldmeier relationship between the maximal area $A_0$ and lifetime 
$T$ of a spot group
\[ A_0=\DGW T \qquad \DGW\simeq 10 \]
is also demonstrated for individual sunspots. No supporting evidence is found
for a supposed supergranular ``quantization'' of sunspot areas. 

The \it robustness\/ \rm of these findings is worth stressing. All the
subtleties of the data processing described in Section 3 will only lead to
slight quantitative modifications in the correlations which are already present
in the raw data. 

Comparing these results with the model predictions of Table I we find that the
turbulent erosion model is compatible with all the observational data while all
the other models are excluded. 
The turbulent erosion model (\citeNP{Petrovay+FMI:erosion}) assumes a strict 
cylindrical geometry with no 
dependence of the field on the coordinate along the tube. This is clearly 
rather far from being a good representation of real sunspots. Yet, the fact that 
the models are able to predict many of the essential features of sunspots, 
including the spontaneous formation of a current sheet around the spot and the 
qualitative and quantitative characteristics of the decay, suggests that the 
model may correctly grasp the essential underlying physics. 

In this paper we have dealt with the \it mean\/ \rm relationships governing the
decay of sunspots. In each case however a significant physical scatter is also 
present in the data. While this scatter is apparently ``random'',
it is already known that the decay properties of sunspots may correlate with 
their other physical characteristics. Specifically, a higher decay rate has
been found to be associated with an irregular shape
(\citeNP{Robinson+Boice:decay}), 
$f$\/-type polarity (\citeNP{RGO}), more bright structures in the umbra
(\citeNP{Zwaan:ARAA}), or higher proper motion (\citeNP{Howard:decay}). The
problem of random and systematic deviations from the mean laws will be treated 
in the following papers of this series.

\acknowledgements 
Valuable comments by Manolo V\'azquez are gratefully acknowledged.  
This work was funded in part by the DGICYT project
no.~91-0530, the DGES project no.~95-0028, 
and by the OTKA under grants no.\ F012817 and T17325.

%\appendix
%\section{dg}

%\bibliography{kris0}
%\bibliographystyle{solphys}

\end{document}